\documentclass[11pt]{article}
\usepackage[textwidth=15.2cm,textheight=22cm]{geometry}
\usepackage{amsmath,amssymb}
\usepackage{latexsym}
\usepackage{multicol}
\usepackage{graphicx}
\usepackage{bigints}
\usepackage{enumitem}
\usepackage{hyperref}
\usepackage{bm}

\hypersetup{
    colorlinks=true,
    linkcolor=blue,
    urlcolor=blue,
}
\usepackage{bm}
\allowdisplaybreaks[1]

\newcommand{\del}{\partial}
\newcommand{\be}{\begin{equation}}
\newcommand{\ee}{\end{equation}}
\newcommand{\ba}{\begin{eqnarray}}
\newcommand{\ea}{\end{eqnarray}}
\newcommand{\bdm}{\begin{displaymath}}
\newcommand{\edm}{\end{displaymath}}

\newcommand{\rom}[1]{\uppercase\expandafter{\romannumeral #1\relax}}

\def\k{\kappa}

\def\ba{\bar A}

\def\beq{\begin{equation}}
\def\eeq{\end{equation}}
\newcommand{\half}{\frac{1}{2}}
\newcommand{\nn}{\nonumber}

\newcommand{\ndt}{\noindent}

\def\bea{\begin{eqnarray}}
\def\eea{\end{eqnarray}}
\def\beas{\begin{eqnarray*}}
\def\eeas{\end{eqnarray*}}
\def\sla{\raise.15ex\hbox{$/$}\kern-.57em}

\def\spa#1.#2{\left\langle#1\,#2\right\rangle}
\def\spb#1.#2{\left[#1\,#2\right]}

\begin{document}

\begin{titlepage}
\begin{flushright}    
{\small $\,$}
\end{flushright}
\vskip 1cm
\centerline{\Large{\bf{Higher derivative quartic vertex }}}
\vskip 0.5cm
\centerline{\Large{\bf{of}}}
\vskip 0.5cm
\centerline{\Large{$\bm{f(R)}$ {\bf gravity in light-cone gauge}}}
\vskip 1.5cm
\centerline{Mahendra Mali and S. Shankaranarayanan}
\vskip 1cm
\centerline{\it {Department of Physics, Indian Institute of Technology Bombay}}
\centerline{\it {Mumbai 400076, India}}
\vskip .5cm
\centerline{Email: \href{mailto: m.mali@iitb.ac.in}{m.mali@iitb.ac.in}, \href{mailto: shanki@phy.iitb.ac.in}{shanki@phy.iitb.ac.in}}
\vskip 1.5cm
\centerline{\bf {Abstract}}
\vskip .5cm

\ndt In the recent studies of four-dimensional Einstein-Hilbert action, quite a few interesting results such as the Kawai-Lewellen-Tye (KLT) relations, MHV-Lagrangian and quadratic forms have been reported. These results naturally raise an important question: Whether these results are valid for the modified theories of gravity in 4-dimensions?  In this work, we consider $R+\alpha\,R^2$ gravity and derive the complete quartic interaction vertex in light-cone gauge. We then discuss the implications of the results for KLT relations, and computing MHV amplitudes.
\ndt

\vfill
\end{titlepage}
\vskip .3cm

\section{Introduction}
\vskip .3cm

General relativity has been highly successful in describing both, the weak field regime (solar system) and the strong field regime
(Black-hole, Neutron stars)~\cite{will}. However, quantum theory of gravity based on the Einstein-Hilbert action and
quantum mechanics is not renormalizable~\cite{hode}. In the recent past,  the study of quantum field theory aspects of general relativity and Supergravity has led to a few interesting results.\\[.2cm]
\ndt In the past three decades, the perturbative analysis of Einstein-Hilbert action has revealed few striking results~\cite{ss, bcl, anthei, ABSM}:

\begin{enumerate}[label=(\alph*)]
\item  The Kawai-Lewellen-Tye (KLT) relations, observed in string theory, state that the closed-string amplitudes factorize into open-string amplitudes~\cite{KLT}. In quantum field theory limit --- in the light-cone gauge --- the KLT relations imply that the tree-level gravity amplitudes are the product of tree-level amplitudes of Yang-Mills theory~\cite{anthei}. Although the KLT 
relations are valid only at the tree-level, they have been used to obtain the quantum loops using unitarity based methods. In principle, the complete (order by order) S-matrix of gravity can be obtained using this method~\cite{kltbern}. Besides, the tree-level cubic amplitude of light-cone higher spin theories when expressed in spinor-helicity formalism manifest KLT-like relations, thereby extending this relation beyond Yang-Mills (spin-1) and gravity (spin-2) system~\cite{ananths}.

\item The tree-level amplitudes in which only two external legs carry negative helicity are called maximally helicity violating (MHV) amplitudes. These amplitudes take very compact form and factorize with spinor-helicity brackets~\cite{shf}. {In Ref.~\cite{anthei}, authors obtain MHV vertices of gravity through field redefinition and the existence of KLT relations at the level of Lagrangian.}
\item {It has been proposed that the surprising ultra-violet finiteness of $\mathcal N\,=\,8$ supergravity is due to the hidden underlying symmetry principle which is present in pure gravity itself~\cite{bern}. {Usually, symmetry principles stem from simple mathematical structures}. It was shown that the Hamiltonian of pure gravity, in four dimensions, can be written in a simple mathematical form --- quadratic forms of an operator $\mathcal D \bar h$~\cite{ABSM}. Also, it was shown that the Hamiltonian up to quartic order is invariant under residual symmetry transformation whereas $\mathcal D \bar h$ transforms covariantly only up to cubic order.  }
    
\end{enumerate}

\ndt On the other hand, the interaction of quantum matter field with the background curvature indicates that the higher order curvature invariants need to be added to the Einstein-Hilbert action~\cite{birrell, utiya}. The cosmological observations of the late-time cosmic acceleration also suggest that the gravity may be modified at the largest possible scales~\cite{review,reviewfR}. In this work, we consider one of the simplest extension of general relativity --- $f(R)$ gravity:
\be 
S\,=\, \bigintsss\,d^4\,x\, \sqrt{-g}\, f(R),
\ee
where $f(R)$ is a power series defined by 
\be
f(R)\,=\,\Sigma_{n=1}^{\infty}\,a_{n}\,R^n \, ,
\ee
where $R$ is the scalar curvature and $a_n$ represents the coefficient of the $n$th term. Amongst modified theories, $f(R)$ gravity is special as it does not suffer from Ostrogradsky instability (no ghost fields) and thus avoids violation of unitarity~\cite{woodard}.\\[.2cm]
\ndt The above mentioned interesting results, in general relativity, motivate us to ask the following questions: Which of the above results is unique to general relativity and which will continue to hold for modified theories of gravity like $f(R)$? Specifically, will the KLT relations be valid for $f(R)$ theories of gravity? In $f(R)$ theories, what happens to the field redefinition and consequently the MHV vertices? Do these quadratic forms exist with the higher derivative interaction terms? How does the Hamiltonian and the quadratic forms transform under the residual symmetry transformation?\\[.2cm]
\ndt In a nutshell, this research program aims to extend the above mentioned results to $f(R)$ gravity. In a similar program, authors have computed the Einstein-Hilbert action in light-cone gauge on maximally symmetric spacetimes with the non-vanishing cosmological constant, i.e., curved backgrounds~\cite{curve}. \\[.2cm]
\ndt Hence, keeping this in mind, the present study is to set the bare-bone of the formalism to obtain fourth order metric derivative correction at quartic order.  To keep the calculations transparent, we first consider $R + \alpha R^2$ model and then extend the results to general $f(R)$. \\[.2cm]
\ndt The organization of the paper is as follows. In Sec.~2, we briefly review the $R+\alpha R^2$ gravity, and field content of the theory. We begin Sec.~3 with the light-cone gauge formulation and discuss the need for a new gauge choice. We expand the action --- up to quartic order including higher derivative interaction vertex and also obtain action in helicity basis (see Sec.~3.3). In Sec.~4, we generalize the results for general $f(R)$-gravity. In the final section, we briefly discuss the implications of the results.\\[.2cm]
In this work, the 4-dimensional metric signature we adopt is
$(-,+,+,+)$, and $\kappa^2 = 8 \pi G$. 

\section{$R\,+\,\alpha\,R^2$ - Action }
\vskip .2cm
We begin with the following action
\be \label{eq:Action} S\,=\,\bigintsss\, d^4\,x\,\, \sqrt{-\,g}\,\,
\bigg[\frac{1}{2\,\k^2}\,R\,+\, \frac{\alpha}{4}\, R^2\,\bigg],\\[.2cm]\
\ee
\ndt where $g$ is the determinant of space-time metric. It is important to note that the quadratic term will contribute when
$\alpha R^2$ is comparable with the Einstein-Hilbert term or in the small length scales. The above action leads to a consistent inflationary model of the early Universe~\cite{staro80}.\\[.2cm]
\ndt The field equations corresponding to the above action \eqref{eq:Action} is
\begin{eqnarray}
\label{eq:eom}
& & \Sigma_{\mu\nu}\,\equiv \,R_{\mu\nu}\,-\,\frac{1}{2}\, g_{\mu\nu}\,R\,+\,\alpha \k^2\,\bigg[R\,R_{\mu\nu}\,-\,\frac{1}{4}\,g_{\mu\nu}\,R^2\,+\,\nabla_\mu\, \nabla_\nu\,R\,-\,g_{\mu\nu}\,\Box\,R\,\bigg]\,=\,0.
\end{eqnarray}
\ndt We write the above field equations as
\be
G_{\mu\nu}\,=\,\kappa^2T^{\text{eff}}_{\mu\nu} \label{eq:effEins},
\ee
\ndt where 
\be
T^{\text{eff}}_{\mu\nu}\,=\, \alpha \left[g_{\mu\nu}\Box R - \nabla_{\mu} \nabla_{\nu} R + \frac{1}{4}g_{\mu\nu}R^{2} - RR_{\mu\nu}\right],\label{eq:Teff}
\ee
\ndt where $\nabla_\mu$ is the covariant derivative and 
$\Box\,=\,\nabla_\mu\,\nabla^\mu$ is the d'Alembartian operator.

\subsection{Field content}
\vskip .2cm 
\ndt The addition of higher curvature terms to the Einstein-Hilbert action preserves diffeomorphism invariance. However, $f(R)$ theory has an extra scalar degree of freedom~\cite{KS78}. In fact, through conformal transformation, it can be shown that the quadratic theory ($R + \alpha R^2$) is equivalent to the Einstein-Hilbert action plus a scalar field (minimally coupled scalar-tensor theory of gravity)~\cite{han72}. The above equivalence is valid for general $f(R)$ theory~\cite{KM89,reviewfR}. By making use of
Newman-Penrose formalism~\cite{new62}, it was found that the scalar field induces the longitudinal polarization and the authors claimed that there are four degrees of freedom~\cite{cor07}. Subsequently, this claim was supported by arguing that the traceless condition cannot be implemented~\cite{kaus16}.\\[.2cm]
\ndt However, later it was pointed out that there is no problem in implementing the transverse-traceless condition and there are only three degrees of freedom~\cite{myun16}. Liang {\it et al.} showed that there are three physical degrees of freedom in $f(R)$ gravity, i. e., plus `$+$', cross `$\times$', and a mix of transverse and longitudinal polarization excited by the massive scalar degree of freedom~\cite{lian17}.

\section{Perturbative action}
\vskip .2cm
\ndt For the Minkowski spacetime metric  with signature $(-, +, +, +)$, the light-cone coordinates are \\
\be x^{\pm}\,=\,\frac{x^0\,\pm\,x^3}{\sqrt 2},
\qquad \partial_{\pm}\,=\,\frac{1}{\sqrt 2}\,(\del_0\,\pm\,\del_3).
\ee 
\ndt We choose $x^+$ as the light-cone time coordinate and
$-\, i\,\del_+$ as the light-cone Hamiltonian. Now $\del_-$ is the spatial
derivative and its inverse $\frac{1}{\del_-}$ (an operator unique to
light-cone field theory) is an integration and defined using the
prescription given in Ref.~\cite{mandel}. Finally, the components
of a vector are $A^\mu\,=\,(A^+, A^-, A^i),$ where $i\,=\,1,2$.\\[.2cm]
\ndt In the covariant formulation, the perturbation of the metric about the
Minkowski background is
\be
g_{\mu\nu}\,=\,\eta_{\mu\nu}\,+\,h_{\mu\nu} \, .
\ee
Using $g_{\mu\nu}\,g^{\nu\rho}\,=\,\delta_\mu^{\,\,\,\,\rho}$, the
inverse of the metric $g_{\mu\nu}$ is given by
\be
g^{\mu\nu}\,=\,\eta^{\mu\nu}\,-\,h^{\mu\nu}\,+\,h^{\mu\rho}\,h_\rho^{\,\,\,\,\nu}\,-\,h^{\mu\rho}\,h_{\rho\sigma}\,h^{\nu\sigma}\,+\,h^{\mu\rho}\,h_\rho^{\,\,\,\,\sigma}\,h^{\nu\alpha}\,h_{\alpha\sigma}\,+\, \dots.\\[.2cm]\
\ee
\ndt Note that the indices are raised and lowered using $\eta_{\mu\nu}=\text{diag}(-, +, +, +)$, i. e., $h^{\mu\nu}\,=\,\eta^{\mu\alpha}\,\eta^{\nu\beta}\,h_{\alpha\beta}$.

\subsection{Gauge choices and constraint relations}
\vskip .2cm
\ndt The invariance of the Einstein-Hilbert action under general coordinate transformation allows four gauge choices in four dimensions. By solving the constraint equations of motion, it is possible to eliminate all the other unphysical degrees of freedom. This procedure generates higher order interaction terms, a characteristic feature of gauge field theory formulations in light-cone gauge.\\[.2cm]
\ndt The perturbative Einstein-Hilbert action in light-cone gauge has been extensively studied~\cite{ss, bcl}. In the case of the Einstein-Hilbert action, the usual gauge choices in light-cone coordinates are
$h_{- \mu}\,=\,0$~\cite{ss, bcl}. However, these gauge choices are not
useful, for $f(R)$, due to the following reasons:
\begin{enumerate}
\item In the case of general relativity, $R$ satisfies an algebraic equation $R\,=\,0$. However, in the case of $R\,+\,\alpha\,R^2$ gravity, $R$ satisfies an independent wave-equation~\cite{reviewfR}. Taking the trace of the equations of motion~\eqref{eq:eom}, we get:
    \begin{equation}
          \label{eq:trace}
    \Box R - \frac{1}{3 \alpha \kappa^2} R = 0.
    \end{equation}
The above wave equation should not come as a surprise because $R\,+\alpha\,R^2$ gravity is a higher derivative theory and contains the fourth derivative of the metric. The advantage of this theory is that the fourth order derivative separates into two second-order equations of motion --- Effective Einstein's equations~\eqref{eq:effEins} and the wave equation of $R$. To obtain higher order vertices, we need to take into account the above wave equation. 
\item It is important to note that the trace equation (\ref{eq:trace}) does not vanish when we consider metric perturbations about the flat background. In other words, the perturbed Ricci scalar is like a propagating scalar field in flat spacetime. Thus, the presence of propagating scalar curvature in the field equation~\eqref{eq:eom} makes the constraint equations unsolvable. Consequently, the redundant degrees of freedom cannot be integrated out in terms of physical fields.
\end{enumerate}
\ndt Therefore, in order to make the constraint equations solvable, we make the following gauge 
choices (like in Ref.~\cite{gair}):
\be \label{eq:gc} g_{- -}\,=\,0,\quad g_{- i}\,=\,0, \quad g_{+
  -}\,=\, -\,1\,-\,a\, R, 
\ee 
\ndt or equivalently, $h_{- \mu}\,=\, a\,R\,\eta_{- \mu}$. Here, $ R$ is the Ricci scalar, and $a$ is an unknown parameter which will be determined by the constraint equations. In equation~\eqref{eq:gc}, we have chosen the propagating scalar curvature ($\phi\,=\, R)$ as metric component by setting $h_{+ -}\,=\,-\, a\, R$. As a result of this gauge choice, the Einstein-Hilbert action will contribute to higher derivative interaction vertices (see the Appendix).\\[.2cm]
\ndt Now we linearly expand constraint components of the equations of motion $\Sigma_{\mu\nu}\,=\,0$ with Ricci tensor defined as
$R_{\mu\nu}\,=\,\del_\nu\,\Gamma^\rho_{\rho\mu}\,-\,\del_\rho\,\Gamma^\rho_{\mu\nu}\,+\Gamma^\rho_{\nu\sigma}\,\Gamma^\sigma_{\rho\mu}\,-\,\Gamma^\rho_{\rho\sigma}\,\Gamma^\sigma_{\mu\nu}$. Using~\eqref{eq:gc}, $\Sigma_{- -}\,=\,0$ yields
\be \del_-^2\,h^\mu_{\,\,\,\mu}\,+\,
(-\,2\,a\,+\,\alpha\,\k^2)\,\del_-^2\, R\,=\,0. \\[.2cm]\ 
\ee
\ndt For $a\,=\,\frac{\alpha}{2}\,\k^2$, the above equation leads to \\
\be \del_-^2\,h^\mu_{\,\,\,\mu}\,=\,0. \\[.2cm]\ \ee 
\ndt For an arbitrary function $h^\mu_{\,\,\,\mu}$, the above relation implies that $h_{\mu\nu}$ is traceless $(h^\mu_{\,\,\,\mu}\,=\,0)$~\cite{BBK}. The other constraint relations yield 
\be \label{eq:hpi} \Sigma_{- i}\,=\,0
\quad \Rightarrow \quad
h_{+i}\,=\,\frac{\del^j}{\del_-}\,h_{ij}\,-\,\frac{\alpha}{2}\,\k^2\,\frac{\del_i}{\del_-}\,R,
\ee 
\ndt and 
\be \label{eq:hpp} \Sigma_{+ -}\,=\,0 \quad \Rightarrow \quad
h_{++}\,=\,
\frac{\del^i\,\del^j}{\del_-^2}\,h_{ij}\,+\,\alpha\,\k^2\,\frac{\del_+}{\del_-}\,R\,-\,\alpha\,\k^2\,\frac{\del^i\,\del_i}{\del_-^2}\,
R.  
\ee
\ndt Observe that the two unphysical degrees of freedom, $h_{+ i}$ and $h_{++}$, can be written in terms of physical fields $h_{ij}$. The inverse of $h_{+ i}$ and $h_{++}$ are related through the following 
relations.
\be h^{+ -}\,=\,h_{+ -}\,=\,-\,\frac{\alpha}{2}\,\k^2\,R; \qquad h^{-
  i}\,=\,-\,\eta^{i j}\,h_{+ j}
\ee
Note that we consider only first-order terms while eliminating the unphysical degrees of freedom and also to obtain the above inverse
relation. Further, we redefine the field 
\be h_{ij}\quad \rightarrow \quad h'_{ij}\,=\,h_{ij}
\,-\,\frac{\alpha}{2}\,\k^2\,\delta_{ij}\,R,
\ee
and observe that the new field $h'_{ij}$ is traceless,
ie. $h'_{22}\,=\,-\,h'_{11}$. For further calculations we drop the
prime sign from $h'_{ij}$. Using the equations
\eqref{eq:gc},~\eqref{eq:hpi} and~\eqref{eq:hpp}, we obtain
\be \sqrt{-\,g}\,=\, (1\,+\,h_{+
  -})\,\sqrt{1\,-\,\frac{h_{ij}\,h_{ij}}{2}}\,=\,1\,+\,h_{+
  -}\,-\,\frac{h_{ij}\,h_{ij}}{4}+\cdots \, .  \ee
\ndt We rescale the field $h$ as \\
\be \label{eq:scale} h\,\quad \rightarrow \quad\,
\frac{h}{\sqrt{2}\,\k}.\\[.2cm]\ 
\ee
In the rest of this section, we expand the action~\eqref{eq:Action} ---  up to quartic order including higher-derivative interaction vertex ---  in the real and helicity basis.

\subsection{Results in the real basis}
\label{sec:results1}
\vskip .2cm
\ndt The action at $\mathcal{O}(h^2)$ is 
\be
S_2\,=\,\bigintsss\, d^4\,x\,\,\, \mathcal L_2 ,
\ee 
\ndt with
\be \label{eq:kine}
\mathcal L_2\,=\, \frac{1}{2}\, h_{ij}\,\del_+\,\del_-\,h_{ij}\,-\,\frac{1}{4}\, h_{ij}\,\del_k\,\del_k\,h_{ij}.
\ee 
\ndt Action at $\mathcal{O}(h^3)$ reads \\
\be 
S_3\,=\,\bigintsss\, d^4\,x\,\,\, \mathcal L_3 ,
\ee
\ndt where 
\bea \label{eq:cubic}
\mathcal L_3\,&\!\!\!\!\!\!=\!\!\!\!\!\!&\, \sqrt{2}\,\k\, {\Bigg\{} -\, h_{kl}\, \del_- h_{li}\,\frac{\del_k\,\del_j}{\del_-}\,h_{ij}\,+\,\frac{1}{2}\,\frac{\del_j}{\del_-}\,h_{ij}\,\del_i h_{kl}\,\del_- h_{kl}\,-\,\frac{1}{4}\,\frac{\del_i\,\del_j}{\del^2_-}\,h_{ij}\,\del_-h_{kl}\,\del_-h_{kl}\nn\\
&&+\,\frac{1}{2}\,h_{il}\,\del_j h_{ij}\,\del_k h_{kl}\,-\,\frac{1}{2}h_{jk}\,\del_i h_{lj}\,\del_l h_{ik}\,-\,\frac{1}{2}\,h_{jk}\,\del_i h_{lj}\,\del_i h_{kl}\,-\,\frac{1}{4}\,h_{ij}\,\del_i h_{kl}\,\del_j h_{kl}\nn\\
&&+\, h_{jk}\,\del_j h_{kl}\,\del_i h_{il} {\Bigg \}}.
\eea
\ndt We see that the cubic vertex derived purely from the Einstein-Hilbert action is free from time derivatives as derived earlier in Ref.~\cite{ss}. Note that the $\alpha\,R^2$ term in the action \eqref{eq:Action} does not contribute to the cubic vertex. This becomes clear from the following arguments.
\bea
S [\text {at}\,\,\mathcal O(h^3)]\, &\sim& \, [R \,\,\text{for}\,\, \eta_{\mu\nu} ]\,\times\, [R\,\, \text {at}\,\, \mathcal O(h^3)]\nn\\[.3cm]
S [\text {at}\,\, \mathcal O(h^3)]\, &\sim& \, [R\,\, \text{at}\,\, \mathcal O (h)]\,\times [R \,\, \text{at}\,\, \mathcal O(h^2)]
\eea
Since, $R \,\,\text{for}\,\, \eta_{\mu\nu}\,(\text{say ---} \bar R)=\,0$  and gauge fixed $R\,\, \text{at}\,\, \mathcal O (h)\,=\,0$. Therefore the contribution of $\alpha\,R^2$ at $\mathcal O(h^3)$ to the Lagrangian vanishes. \\[.2cm]
\ndt The quartic vertex, from the Einstein-Hilbert term of the action, reads as
\bea \label{eq:quart}
\mathcal L_4^{(EH)} \,&\!\!\!\!\!\!=\!\!\!\!\!\!&\, 2\,\k^2{\Bigg\{} -\frac{1}{2}\,h_{ik}\,h_{jl}\,\del_+h_{kl}\,\del_-h_{ij}\,-\,\frac{3}{2}\,h_{jl}\,h_{kl}\,\del_+h_{ik}\,\del_-h_{ij}\,+\,\frac{1}{2}\,h_{jk}\,h_{kl}\,\del_-h_{li}\,\del_+h_{ij}\,\nn\\
&&+\,\frac{1}{2}\,h_{il}\frac{\del_j\del_m}{\del_-^2}h_{jm}\,\del_-h_{kl}\,\del_-h_{ki}\,-\,h_{jk}\,h_{kl}\,\frac{\del_p}{\del_-}\,h_{lp}\,\del_i\del_-h_{ij}\,+\,h_{jk}\,h_{ik}\,\del_ph_{lp}\,\del_lh_{ij}\nn\\
&&-\,h_{jk}\frac{\del_p}{\del_-}h_{lp}\,\del_-h_{ij}\,\del_ih_{lk}\,-\,h_{jk}\,\frac{\del_p}{\del_-}h_{kp}\,\del_-h_{il}\,\del_ih_{jl}\,-\,h_{ik}\,\frac{\del_p}{\del_-}h_{lp}\,\del_-h_{jk}\,\del_l h_{ij}\nn\\
&&-\,h_{jk}\,h_{ik}\,\del_-h_{ij}\,\frac{\del_p\,\del_l}{\del_-}h_{lp}\,+\,2\,h_{jk}\,h_{ik}\,\frac{\del_j\,\del_p}{\del_-}h_{lp}\,\del_-h_{il}\,+h_{jk}\,h_{kl}\,h_{ij}\,\del_i\del_p h_{lp}\nn\\
&&-\,\frac{1}{2}\,h_{jk}\,\frac{\del_p}{\del_-}h_{kp}\,\del_j h_{il}\,\del_-h_{il}\,+\,h_{ik}\,\frac{\del_p}{\del_-}h_{lp}\,\frac{\del_j}{\del_-} h_{kj}\,\del_-^2 h_{il}\,+\,\frac{1}{2}\,h_{kl}\,h_{il}\,\del_j h_{ij}\,\del_p h_{kp}\nn\\
&&+\,h_{kl}\,\frac{\del_j}{\del_-}h_{ij}\,\del_p h_{kp}\,\del_-h_{il}\,+\,\frac{1}{2}\,\frac{\del_p}{\del_-} h_{kp}\,\frac{\del_j}{\del_-} h_{lj}\,\del_-h_{il}\,\del_- h_{ik}\,+\,\frac{1}{4}\,\frac{\del_j}{\del_-} h_{kj}\,\frac{\del_p}{\del_-} h_{kp}\,\del_-h_{il}\,\del_- h_{il}\nn\\
&&+\,h_{kn}\,h_{lj}\,h_{nj}\,\del_i\del_k h_{il}\,+\,\half\,h_{lm}\, h_{km}\,\del_i h_{jl}\, \del_j h_{ik}\,+\,h_{kl}\,h_{jn}\,\del_i h_{ln}\,\del_k h_{ij}\,+\frac{1}{2}\,h_{kj}\, h_{ln}\, \del_k h_{ni}\,\del_l h_{ij}\nn\\
&&-\,\frac{1}{4}\,h_{kn}\,h_{lj}\,\del_i h_{kl}\,\del_i h_{nj}\,-\,\half\,h_{kl}\,h_{nj}\,\del_k h_{in}\, \del_l h_{ij}\,-\,\frac{1}{4}\, h_{lm}\, h_{km}\,\del_k h_{ij}\,\del_l h_{ij}\,+\frac{1}{4}\,h_{kl}\,h_{nj}\,\del_i h_{kl}\,\del_i h_{nj}\nn\\
&&+\,h_{kl}\,h_{nj}\,\del_k h_{ij}\,\del_l h_{ni}\,+\,\half\,h_{jn}\,h_{kn}\,\del_i h_{lj}\,\del_i h_{kl}\,+\,\half\, h_{lj}\,h_{kn}\,\del_i h_{jn}\,\del_i h_{kl}\,-\,h_{kj}\,h_{ln}\,\del_l h_{ij}\,\del_k h_{in}\nn\\
&&-\,h_{ij}\,h_{ln}\, \del_l h_{kj}\,\del_k h_{in}\,+\,\half\,h_{kj}\, h_{ij}\,\del_j h_{ln}\,\del_k h_{ln}\,+\,2\,h_{il}\,h_{jk}\,\frac{\del_j\,\del_p}{\del_-}h_{lp}\,\del_-h_{ik}\,+\,h_{jk}\frac{\del_p}{\del_-}h_{lp}\,\del_j h_{il}\,\del_-h_{ik}\nn\\
&&+\,\half\,h_{kl}\,h_{ij}\,\del_-h_{ij}\,\del_+h_{kl}\,+\,\frac{1}{8}\,h_{kl}\,h_{kl}\,\del_-h_{ij}\,\del_+h_{ij}\,+\,\frac{1}{4}\,h_{kl}\,h_{kl}\,\del_-h_{ij}\,\frac{\del_i\,\del_p}{\del_-}h_{jp}\,+\,\frac{1}{4}\,h_{kl}\,h_{kl}\,h_{ij}\,\del_p\del_p h_{ij}\nn\\
&&-\,\frac{1}{8}\,h_{kl}\,h_{kl}\,\del_i h_{ij}\,\del_p
h_{jp}\,-\,\frac{1}{8}\, h_{kl}\,h_{kl}\,\del_i h_{pj}\,\del_p
h_{ij}\,+\,\frac{3}{16}\,h_{kl}\,h_{kl}\,\del_i h_{jp}\,\del_i
h_{jp}{\Bigg\}}. \eea 
\ndt Due to the gauge choice (\ref{eq:gc}), the Einstein-Hilbert action contributes to the higher-order interaction vertices. (See also Appendix.) The higher-derivative correction to the quartic order is
\bea \label{eq:hdq}
\mathcal L_4^{\alpha} \,&\!\!\!\!\!\!=\!\!\!\!\!&\, 2\,\alpha\,\k^4\,{\Bigg\{}\,\frac{1}{4} R_{(2)}\,\del_-h_{ij}\,\del_+ h_{ij}\,-\,\frac{1}{4}\,\frac{\del_+}{\del_-} R_{(2)}\,\del_- h_{kl}\,\del_- h_{kl}\,+\,h_{kl}\,\del_- h_{li}\,\frac{\del_k\,\del_i}{\del_-} R_{(2)}\nn\\
&&-\,\frac{1}{2}\,\frac{\del_i}{\del_-} R_{(2)}\,\del_i h_{kl}\,\del_- h_{kl}\,-\half\,h_{ij}\,\del_j R_{(2)}\,\del_k h_{ik}\,+\,\frac{3}{8}\,\frac{\del_i\,\del_i}{\del^2_-} R_{(2)}\,\del_-h_{kl}\,\del_- h_{kl}\nn\\
&&+\,\half \, R_{(2)}\frac{\del_i\del_k}{\del_-}h_{jk}\,\del_-h_{ij}\,-\,\frac{1}{4}\, R_{(2)}\,\del_j h_{ij}\,\del_k h_{ik}\,+\frac{1}{4}\, R_{(2)}\,\del_i h_{kj}\,\del_k h_{ij}\,+\,\half h_{jk}\,\del_i  R_{(2)}\,\del_i h_{jk}\nn\\
&&+\,\frac{3}{8} R_{(2)}\,\del_i h_{kl}\,\del_i h_{kl}\,+\,\half
h_{ij} R_{(2)}\,\del_i \del_k h_{jk}\,-\,\frac{1}{4}
R^2_{(2)}{\Bigg\}},
\eea 
\vskip .2cm 
\ndt where 
\bea
R_{(2)}&=&\frac{3}{2} \del_-\, h_{ij}\,\del_+ h_{ij}\,+\,2\,h_{ij}\del_-\del_+h_{ij}\,-\,\del_- h_{ij}\,\frac{\del_i \del_k}{\del_-}\,h_{jk}\,-\,h_{ij}\,\del_k \del_k h_{ij}\,+\,\half\,\del_i h_{ij}\,\del_k h_{jk}\nn\\
&&+\,\frac{1}{2}\,\del_i h_{kj}\,\del_k h_{ij}\,-\,\frac{3}{4}\,\del_i
h_{jk}\,\del_i h_{jk}.\nn \eea
\ndt It is important to note that we have dropped the boundary terms while obtaining the above results. In the previous studies of light-cone gravity, redundant degrees of freedom are eliminated by expressing them in terms of higher order (in fact infinite order) function of the physical field
$h_{ij}$~\cite{ss, bcl}. This is achieved by solving the constraint equations. This procedure introduces interaction terms which are not
obtained in the usual covariant formulation. For instance, this methodology has been utilized to obtain the enormous quintic interaction vertex of the Einstein-Hilbert action~\cite{ananthq}. In this paper, we obtain the unphysical fields in terms of physical fields only up to linear order, therefore, equation~\eqref{eq:quart} does not contain all the interaction terms. However, a suitable field redefinition should suffice to link the interaction terms in equation~\eqref{eq:quart} with that obtained in Refs.~\cite{ss, bcl}. (Note that the cubic term matches with Ref.~\cite{ss,bcl} up to a factor of $2$.) {On the contrary, we require only the first order perturbations to compute the fourth order derivative correction terms to the Lagrangian at $\mathcal O(h^4)$. } Therefore, the fourth order
derivative quartic vertex in equation~\eqref{eq:hdq} is the complete expression.

\subsection{Results in the helicity basis} 
\vskip.2cm
\ndt In this section, we compute the perturbative action in helicity eigenbasis. As it is known, to specify the helicity states of a 
particle with spin $s$, it is not necessary to have the equation of motion for such a particle, it is enough to know that the equation
exists. For the action~\eqref{eq:Action}, this amounts to the fact that we do not need the explicit equation of motion of $R$, it is sufficient to know that an equation of motion of $R$ exists.\\[.2cm]
\ndt Here we list all the results of Sec.~(\ref{sec:results1}) in helicity eigenbasis. We obtain this by transforming the transverse coordinates and their derivatives as 
\bea
x\,&=&\,\frac{1}{\sqrt 2}\,(x^1\,+\,i\,x^2), \qquad \bar{\del}\, \equiv\,\frac{\del}{\del\,x}\,=\,\frac{1}{\sqrt 2}\,(\del_1\,-\,i\,\del_2),\nn\\[.2cm]
\bar{x}\,&=&\,\frac{1}{\sqrt 2}\,(x^1\,-\,i\,x^2), \qquad \del\,
\equiv\,\frac{\del}{\del\,\bar x}\,=\,\frac{1}{\sqrt
  2}\,(\del_1\,+\,i\,\del_2).  
\eea 
The fields transform as 
\be
h\,\equiv\,\frac{1}{\sqrt 2}\,(h_{11}\,+\,i\,h_{12}), \qquad \bar
h\,\equiv\,\frac{1}{\sqrt 2}\,(h_{11}\,+\,i\,h_{12}),
\ee 
where $h$ and $\bar h$ represent `+2' and `-2' helicities of graviton respectively. The Lagrangian in the equation~\eqref{eq:Action}, up to cubic order, reads 
\bea
\mathcal L\,&\!\!\!\!\!\!=\!\!\!\!\!& \,- h\,\Box \bar h\,+\,2\kappa {\Bigg\{} \bar h \del h\,\del \bar h\,-2\,\bar h \del_-h\,\frac{\del\del}{\del_-}\bar h\,+\,\frac{\del}{\del_-}\bar h(\del\bar h\,\del_-h\,+\,\del h\,\del_-\bar h)\nn\\
&&-\,\frac{\del\del}{\del_-^2}\bar h\,\del_-h\,\del_-\bar
h{\Bigg\}}\,+\,c.c. .\eea 
\ndt The quartic vertex reads \bea
\mathcal L_4^{(EH)}\,&\!\!\!\!\!\!=\!\!\!\!\!\!&\, 2\,\k^2{\Bigg\{} 2\,h \bigg[\frac{\del}{\del_-}\bar h\,\del_-\bar h\,\bar{\del}h\,+\,2\frac{\bar \del}{\del_-}h\,\frac{\del}{\del_-}\bar h\,\del_-^2\bar h\,-\,3\frac{\del}{\del_-}\bar h\,\del_-h\,\bar{\del}\bar h\bigg]\,+\, 2\bar h{\bigg[}\frac{\bar \del}{\del_-}h\,\del_-h\,\del \bar h\nn\\[.2cm]
&& +\,2\frac{\bar \del}{\del_-}h\,\frac{\del}{\del_-}\bar h\,\del_-^2 h\,-\,3\frac{\bar \del}{\del_-}h\del_-\bar h\,\del h{\bigg]}\,-\,h\bar h\big(\del_+ h\,\del_-\bar h\,+\,\del_+\bar h\,\del_- h\big)\,+\,8\frac{\del}{\del_-}\bar h\,\frac{\bar \del}{\del_-}h\,\del_-h\,\del_-\bar h\nn\\[.2cm]
&&+\,h\bar h\bigg[10\bigg(\frac{\del\bar \del}{\del_-}\bar h\del_- h+\,\frac{\del\bar \del}h\,\del_-\bar h\bigg)\,-4\,\bigg(\frac{\bar\del}h\,\del\del_-\bar h\,+\,\frac{\del}{\del_-}\bar h\bar\del \del_-h\bigg)\,+11\,\del h\,\bar\del\bar h\,+7\,\del\bar h\,\bar \del h\bigg]\nn\\[.2cm]
&&+hh\bigg(8\,\frac{\del\bar\del}{\del_-}\bar h\,\del_-\bar
h\,+12\,\bar h\, \del\bar\del\bar h+2\del\bar h\,\bar\del\bar
h\bigg)+\bar h\bar
h\bigg(8\,\frac{\del\bar\del}{\del_-}h\,\del_-h +12 h\,\del\bar\del
h\,+2\,\del h\,\bar\del h\bigg){\Bigg\}}. 
\eea 
\vskip .2cm \ndt Finally, the higher derivative action at quartic order is
\bea \label{eq:hdh}
\mathcal L_4^{\alpha}\,&\!\!\!\!\!\!=\!\!\!\!\!&\, 2\,\alpha\,\k^4\,{\Bigg\{} \frac{1}{2}R_{(2)}\big(\del_-h\,\del_+\bar h\,+\,\del_-\bar h\,\del_+h\big)\,-\,\frac{\del_+}{\del_-} R_{(2)}\,\del_-h\,\del_-\bar h\,+\,2\big(h\,\del_-\bar h+\bar h\,\del_-h\big)\frac{\del\bar\del}{\del_-} R_{(2)}\nn\\[.2cm]
&&-\frac{\del}{\del_-} R_{(2)}\,\big(\bar \del h\,\del_-\bar h\,+\,\bar\del\bar h\,\del_- h\big)-\,\frac{\bar \del}{\del_-} R_{(2)}\big(\del h\,\del_-\bar h\,+\,\del \bar h\,\del_- h\big)\,+\,3\frac{\del\bar\del}{\del_-^2}R_{(2)}\,\del_-h\,\del_-\bar h\nn\\[.2cm]
&&+\, R_{(2)}\frac{\del\bar\del}{\del_-}h\,\del_-\bar h\,+\,
R_{(2)}\,\frac{\del\bar \del}{\del_-}\bar h\,\del_-h\,+2\, R_{(2)}\del
h\,\bar \del \bar h\,+2\, R_{(2)}\,\del\bar h\, \bar \del h \,-\,\frac{1}{4}
R^2_{(2)} {\Bigg\}},
\eea \ndt where \bea
R_{(2)}&\!\!\!\!\!\!=\!\!\!\!\!& 3\big(\del_-h\,\del_+\bar h+ \del_-\bar h\,\del_+h\big)+4\,\big(h\,\del_-\del_+\bar h+\bar h\,\del_-\del_+h\big)-4\,\big(h\,\del\bar \del\bar h\,+\,\bar h\del\bar\del h\big)\nn\\
&&-2\,\bigg(\del_-h\,\frac{\del\bar \del}{\del_-}\bar h\,+\,\del_-\bar
h\, \frac{\del\bar \del}{\del_-}h\bigg)\,+\,\bar \del h\,\del\bar
h+\,\del h\,\bar \del\bar h.\nn \eea

\section{Generalization to generic $f(R)$ gravity}
\vskip .2cm
\ndt The action of $f(R)$ theory of gravity in vacuum  is 
\be 
S\,=\, \bigintsss\,d^4\,x\, \sqrt{-g}\, f(R),
\ee
\ndt where 
\be \label{eq:frdef}
f(R)\,=\,\Sigma_{n=1}^{\infty}\,a_{n}\,R^n \quad 
\mbox{with} \quad a_1\,=\,\frac{1}{2\kappa^2},\,\,\,\, a_2\,=\,\frac{\alpha}{4}.
\ee
The corresponding equation of motion can be written as in 
eq.~\eqref{eq:effEins}, where the effective energy-momentum tensor is 
given by
\be
T^{\text{eff}}_{\mu\nu}\,=\,\frac{2\,\kappa^2}{f^{\prime}(R)}\big\{g_{\mu\nu}\,\Box
f^{\prime}(R)\,-\,\nabla_\mu\nabla_\nu
f^{\prime}(R)\,+\,\frac{g_{\mu\nu}}{2}[f(R)\,-\,R\,f^{\prime}(R)]\big\},
\ee
\ndt where $f^\prime(R)\,=\,\del f(R)/\del R$ is\\
\be 
f^\prime(R)\,=\, \Sigma_{n=1}^{\infty}\,n\,a_{n}\,R^{(n-1)}.
\ee
The variation of the above equation under perturbation is  \\
\be \label{eq:deltafp}
\delta\, f^\prime(R)\,=\,2\, a_2\,\delta R\,+\,6\,a_3\,\bar R\,\delta R\,+\, \text{higher order terms},
\ee   
\vskip .2cm
\ndt where $\bar R$ is scalar curvature corresponding to the background metric and $\delta R$ is variation in the scalar curvature due to perturbation. For flat spacetime, $\bar R\,=\,0$, therefore only the first term~\eqref{eq:deltafp} survives and all the higher order terms vanish. We substitute $a_2\,=\,\frac{\alpha}{4}$ in~\eqref{eq:deltafp} and get \\
 \be 
\delta f^\prime (R)\,=\, \frac{\alpha}{2}\delta R.
\ee
\ndt About $\bar R\,=\,0$ \cite{Bhattacharyya:2017tyc}, 
the variation of $T^{\text{eff}}_{\mu\nu}$ under perturbation 
$g_{\mu\nu}\, = \, \bar{g}_{\mu\nu}\,+\delta\,g_{\mu\nu}$ yields\\
\be 
\label{eq:delt} \delta T^{\text{eff}}_{\mu\nu}\,=\, 2\,\kappa^2
\big\{ \bar{g}_{\mu\nu}\,\Box \delta f^\prime (R)\,-\,\nabla_\mu \nabla_\nu
\delta f^\prime (R)\big\}.
\ee 
\vskip .2cm
\ndt After substituting $\delta f^\prime (R)$ in equation \eqref{eq:delt}, we observe that it is identical to the $\delta T^{\text{eff}}_{\mu\nu}$ obtained from equation \eqref{eq:Teff}. Therefore, all the results obtained above for $R+\alpha R^2$ gravity are valid for $f(R)$ gravity. Note that the above analysis is valid only when $f(R)$ is a power series of scalar curvature as given in eq.~\eqref{eq:frdef}.\\[.2cm]
\ndt We would like to point to the readers that, we have reported the formalism to compute the higher derivative interaction vertices of $f(R)$ gravity. In this work, we have explicitly obtained the quartic interaction vertex. For the future research, this formalism can be utilized to compute the higher order interaction vertices (higher than quartic) of $f(R)$ gravity. This can be done by incorporating the higher order  terms in the constraint relations~\eqref{eq:hpi}-\eqref{eq:hpp} and extending $R_{(2)} \rightarrow R_{(3)} $ (where $R_{(3)}$ contains terms at order $h^{3}$) and so on.
 
\section{Discussion and conclusions}
\vskip .2cm 
\ndt As expected the final result in equation~\eqref{eq:hdh} involves scalar curvature, which we can write in terms of two polarizations of the gravitational field. Observe that the equation~\eqref{eq:hdh} contains terms, both, linear and quadratic in time derivative: $\del_+$. In Ref.~\cite{bcl}, the authors showed that the linear $\del_+$ dependence of the quartic vertex could be lifted to higher orders using field re-definition. However, due to the quadratic terms in $\del_+$ the method in Ref.~\cite{bcl} is not applicable here. It will be interesting to find the type of field redefinition that would make the quartic derivative correction to the action free from $\del_+$. The removal of $\del_+$ dependence from interaction vertices is necessary for the realization of MHV-amplitudes and to obtain the quadratic forms of the Hamiltonian.\\[.2cm]
\ndt There have been recent studies to obtain the KLT relations for conformal gravity~\cite{JMT}. Specifically, the authors study the scattering amplitudes of conformal gravity theories and explore the double-copy/KLT relations to develop a better understanding of the conformal gravity. However, it will be interesting to explore the KLT relations relating $f(R)$ gravity (without Weyl terms) and four derivative gauge theory scattering amplitudes in spinor helicity formalism~\cite{shf}. The quartic derivative Lagrangian of Yang-Mills theory that we can consider is~\cite{frad}
\be 
\mathcal L\,=\,-\frac{1}{2\,g^2}\, \text{tr} (F_{\mu \nu})^2\, -\,\frac{1}{m^2\,g^2}\,\text{tr}\bigg[ (\nabla_{\alpha}\,F_{\mu\nu})^2\,+\,\gamma\,F_{\mu\nu}\,[F_{\mu\lambda},\,F_{\nu\lambda}]\bigg],
\ee
where $m$ is a dimensionful parameter and $\gamma$ a constant. \\[.2cm]
\ndt However, it is unclear what would be the stringy origin of this KLT-type relation in higher derivative gravity and Yang-Mills theory. \\[.2cm]
\ndt The scalar-tensor theory of gravity and $f(R)$-gravity are conformally equivalent~\cite{han72}. Can we establish a similar equivalence between the two theories at the level of perturbative Lagrangian~\cite{mali}? Also, in a recent study of one-loop
divergences of $f(R)$-gravity and scalar-tensor theory of gravity, it was shown that the classical equivalence between the two breaks by off-shell quantum corrections, however, restored on-shell~\cite{ruf}.\\[.2cm]
\ndt As mentioned above, the $f(R)$ gravity is conformally related to scalar-tensor theory which leads to a consistent inflationary model of the early Universe~\cite{staro80}.  In cosmology, beyond linear perturbation theory, it is difficult to simplify the perturbation equations with gauge-invariant variables alone~\cite{debottam}. Recently, several field theory techniques like BRST are employed to go beyond the linear order~\cite{picon}. The results obtained here is an initial step to use light-cone gauge for further studies of the modified theories gravity. 
\section*{Acknowledgments}
The authors wish to thank U.~Yajnik for discussions. This work is supported by DST-Max Planck Partner Group on Gravity and Cosmology, and IITB-IRCC seed grant. The algebraic calculations partially have been done using Cadabra~\cite{cad} and MM thanks Kasper Peeters for useful suggestions with Cadabra.

\renewcommand{\theequation}{A-\arabic{equation}}
\setcounter{equation}{0}  
\section*{Appendix} 
\vskip 0.5cm

\subsection*{Contributors to higher derivative action at $\mathcal O(h^4)$ }
\vskip 0.5cm
We list below the contributions to equation~\eqref{eq:hdq} from the kinetic and cubic vertices of the Einstein-Hilbert action.\\
\vskip .2 cm
{\bf Contributions from kinetic terms:}
\be \label{eq:kc}
\mathcal L_4\,=\, 2\,\alpha\,\kappa^4 \big\{-\frac{3}{4}\,R_{(2)}^2\big\}
\ee
\vskip .5cm
{\bf Contributions from cubic terms:}
\bea \label{eq:cc}
\mathcal L_4\,&\!\!\!\!\!\!=\!\!\!\!\!&\, 2\,\alpha\,\k^4\,{\Bigg\{}\,\frac{1}{4} R_{(2)}\,\del_-h_{ij}\,\del_+ h_{ij}\,-\,\frac{1}{4}\,\frac{\del_+}{\del_-} R_{(2)}\,\del_- h_{kl}\,\del_- h_{kl}\,+\,h_{kl}\,\del_- h_{li}\,\frac{\del_k\,\del_i}{\del_-} R_{(2)}\nn\\
&&-\,\frac{1}{2}\,\frac{\del_i}{\del_-} R_{(2)}\,\del_i h_{kl}\,\del_- h_{kl}\,-\half\,h_{ij}\,\del_j R_{(2)}\,\del_k h_{ik}\,+\,\frac{3}{8}\,\frac{\del_i\,\del_i}{\del^2_-} R_{(2)}\,\del_-h_{kl}\,\del_- h_{kl}\nn\\
&&+\,\half \, R_{(2)}\frac{\del_i\del_k}{\del_-}h_{jk}\,\del_-h_{ij}\,-\,\frac{1}{4}\, R_{(2)}\,\del_j h_{ij}\,\del_k h_{ik}\,+\frac{1}{4}\, R_{(2)}\,\del_i h_{kj}\,\del_k h_{ij}\,+\,\half h_{jk}\,\del_i  R_{(2)}\,\del_i h_{jk}\nn\\
&&+\,\frac{3}{8} R_{(2)}\,\del_i h_{kl}\,\del_i h_{kl}\,+\,\half h_{ij} R_{(2)}\,\del_i \del_k h_{jk}{\Bigg\}}
\eea
\vskip .2cm
The action in equation~\eqref{eq:hdq} is obtained by adding the equations~\eqref{eq:kc},\eqref{eq:cc} and  second term of equation~\eqref{eq:Action} (after rescaling it as defined in equation~\eqref{eq:scale}).


\begin{thebibliography}{Ref}
\bibitem{will}{C. M. Will, {\it Theory and experiment in gravitational physics}; UK: Univ. Pr. (1993)\\
C. M. Will, {\it The Confrontation between General Relativity and Experiment }, \href{http://dx.doi.org/10.12942/lrr-2014-4}{Living Rev. Rel. {\bf 17, 4} (2014)}.}
\bibitem{hode}{G. 't Hooft, M. Veltman, {\it Ann.\ Inst.\ H.\ Poincare Phys.\ Theor.} {\bf A20} (1974) 69. \\
S.~Deser and P.~van Nieuwenhuizen, \href{http://dx.doi.org/10.1103/PhysRevD.10.401} {{\it Phys.Rev.} {\bf D10} 401-410 (1974)}.} \\
Marc H. Goroff, and Augusto Sagnotti, \href{http://dx.doi.org/10.1016/0550-3213(86)90193-8}{ {\it Nuc. Phys. B}, {\bf 266}, 709-736 (1986)}.
\bibitem{ss}{J. Scherk and J. Schwarz, \href{http://dx.doi.org/10.1007/BF00761962}{ {\it Gen. Rel. Grav.} {\bf 6}, (1975) 537-550}.\\
	M. Goroff and J. Schwarz, \href{http://dx.doi.org/10.1016/0370-2693(83)91630-1}{ {\it Phys. Lett.} {\bf B 127}, (1983) 61-64}.}
\bibitem{bcl}{Ingemar Bengtsson, Martin Cederwall and Olof Lindgren, Goteborg-83-55, (1983).}

\bibitem{anthei}{S. Ananth, S. Theisen, \href{http://dx.doi.org/10.1016/j.physletb.2007.07.003}{ {\it Phys.Lett. B} {\bf 652} (2007) 128-134}.}
\bibitem{ABSM}{S. Ananth, L. Brink, S. Majumdar, M. Mali, N. Shah, \href{http://dx.doi.org/10.1007/JHEP03(2017)169}{ {\it JHEP} {\bf 1703} (2017) 169}.}
\bibitem{KLT}{H. Kawai, D.C. Lewellen and S.H.H. Tye, \href{http://dx.doi.org/10.1016/0550-3213(86)90362-7} { {\it Nucl. Phys. B} {\bf 269} (1986) 1-23}.}
\bibitem{kltbern}{ Z. Bern, J. J. Carrasco, Wei-Ming Chen, H. Johansson, R. Roiban, \href{http://dx.doi.org/10.1103/PhysRevLett.118.181602} {{\it Phys.Rev.Lett.} {\bf 118} (2017) no.18, 181602},\\
Z. Bern, L. J. Dixon, M. Perelstein and J.S. Rozowsky, \href{http://dx.doi.org/10.1016/S0550-3213(99)00029-2} {{\it Nucl. Phys. B} {\bf 546} (1999) 423-479}.}
\bibitem{ananths}{S. Ananth, \href{http://dx.doi.org/10.1007/JHEP11(2012)089}{{\it JHEP} {\bf 1211} (2012) 089}}.

\bibitem{shf}{H. Elvang, Y. Huang, {\sl Scattering Amplitudes in Gauge theory and Gravity}, Cambridge University Press, UK (2015) [\href{http://arxiv.org/abs/arXiv:1308.1697}{ arXiv:1308.1697v2}]\\
L. Dixon, {\it Calculating scattering amplitudes efficiently}, 
in Proceedings {\sl QCD and beyond: TASI-95}, (1996) [\href{http://arxiv.org/abs/hep-ph/9601359}{arXiv:hep-ph/9601359}.]} 

\bibitem{bern}{Z. Bern, J. Carrasco, D. Forde, H. Ita and H. Johansson, \href{http://dx.doi.org/10.1103/PhysRevD.77.025010}{ {\it Phys. Rev.} {\bf D 77} (2008) 025010}.}

\bibitem{birrell}{N. D. Birrell, P. C. W. Davies, {\it Quantum fields in curved space}, Cambridge University Press, Cambridge 1982.}
\bibitem{utiya}{R. Utiyama, B. S. DeWitt, \href{http://dx.doi.org/10.1063/1.1724264}{ {\it J. Math. Phys.} {\bf 3}, 608 (1962)}.}
%
\bibitem{review}
S.~Nojiri and S.~D.~Odintsov, \href{http://dx.doi.org/10.1142/S0219887807001928}{ {\it Int.\ J.\ Geom.\ Meth.\ Mod.\ Phys.}  {\bf 4}, 115 (2007), 115-146},\\
T. Clifton {\it et al.} {\it Modified gravity and Cosmology}, \href{http://dx.doi.org/10.1016/j.physrep.2012.01.001} { {\it Phys.Rept.}{\bf 513} (2012) 1-189}.
\bibitem{reviewfR}
{A. D. Felice, S. Tsujikawa, {\it $f(R)$ Theories }, \href{http://dx.doi.org/10.12942/lrr-2010-3}{ {\it Living Rev. Relativity}, {\bf13}, (2010), 3},\\
T. P. Sotiriou, V. Faraoni {\it $f(R)$ theories of gravity}, \href{http://dx.doi.org/10.1103/RevModPhys.82.451} { {\it Rev.Mod.Phys.} {\bf 82} (2010) 451-497}. }

\bibitem{woodard}{R. P. Woodard, \href{http://dx.doi.org/10.1007/978-3-540-71013-4_14} {{\it Lect. Notes Phys.} {\bf 720} (2007) 403–433}.}
\bibitem{curve}{Y.S. Akshay, S. Ananth, M. Mali, \href{http://dx.doi.org/10.1016/j.nuclphysb.2014.04.015}{ {\it Nucl.Phys. B}{\bf 884} (2014) 66-73}\\
S. Ananth, M. Mali, \href{http://dx.doi.org/10.1016/j.physletb.2015.04.006}{{\it Phys.Lett. B}{\bf 745} (2015) 48-51 }.}






\bibitem{staro80} {A.A. Starobinsky, \href{http://dx.doi.org/10.1016/0370-2693(80)90670-X} {{\it Phys. Lett. B} {\bf 91}, (1980) 99-102}.}
\bibitem{KS78} {K. Stelle,\href{http://dx.doi.org/10.1007/BF00760427}{ {\it Gen. Rel. Grav.} {\bf 9} (1978), 353-371}.}
\bibitem{han72}{J. O'\ Hanlon, \href{http://dx.doi.org/10.1103/PhysRevLett.29.137}{ {\it Phys. Rev. Lett.} {\bf 29}, (1972) 137-138},\\
P. Teyssandier and P. Tourrenc, \href{http://dx.doi.org/10.1063/1.525659} { {\it J. Math. Phys.} {\bf 24}, (1983) 2793}.}



\bibitem{KM89}{K. Maeda, \href{http://dx.doi.org/10.1103/PhysRevD.39.3159}{ {\it Phys. Rev. D} {\bf 39}, (1989) 3159},\\
S. Gottlober, H.-J. Schmidt, A.A. Starobinsky, \href{http://dx.doi.org/10.1088/0264-9381/7/5/018}{{\it Class. Quantum Grav.} {\bf 7}, (1990) 893}.}

\bibitem{new62}{E. Newman, R. Penrose, \href{http://dx.doi.org/10.1063/1.1724257}{{\it J. Math. Phy.} {\bf 3}, (1962) 566-578},\\
D.M. Eardley, D. L. Lee, A. P. Lightman, \href{http://dx.doi.org/10.1103/PhysRevD.8.3308}{{\it Phy. Rev. D} {\bf 8} (1973) 3308-3321}.}
\bibitem{cor07}{C. Corda, \href{http://dx.doi.org/10.1088/1475-7516/2007/04/009}{ {\it JCAP} {\bf 0704}, (2007) 009}.\\
C. Corda, \href{http://dx.doi.org/10.1142/S0217751X08038603}{ {\it Int. J. Mod. Phys. A} {\bf 23}, (2008) 1521}.}

\bibitem{kaus16}{H. R. Kausar, L. Philippoz, and P. Jetzer, \href{http://dx.doi.org/10.1103/PhysRevD.93.124071}{ {\it Phys. Rev. D} {\bf 93}, (2016) no.12, 124071}.}
\bibitem{myun16}{Y. S. Myung, \href{http://dx.doi.org/10.1155/2016/3901734} { {\it Adv. High Energy Phys.} {\bf 2016}, (2016) 3901734}.}
\bibitem{lian17}{D. Liang, Y. Gong, S. Hou, Y. Liu, \href{http://dx.doi.org/10.1103/PhysRevD.95.104034} { {\it Phys. Rev. D} {\bf 95}, (2017) no.10, 104034}.}
\bibitem{mandel}{ S. Mandelstam, \href{http://dx.doi.org/10.1016/0550-3213(83)90179-7}{ {\it Nucl. Phys.} {\bf B 213} (1983) 149-168}.}
\bibitem{gair}{C. P. L. Berry, J. R. Gair, \href{http://dx.doi.org/10.1103/PhysRevD.83.104022}{{\it Phys.Rev. D}{\bf 83} (2011) 104022}.}
\bibitem{BBK}{A.K H. Bengtsson, L. Brink, Sung-Soo Kim, \href{https://doi.org/10.1007/JHEP03(2013)118}{{\it JHEP} {\bf 1303} (2013) 118}.}
\bibitem{ananthq}{S. Ananth, \href{http://dx.doi.org/10.1016/j.physletb.2008.05.035}{\it{Phys.Lett. B} {\bf 664} (2008) 219-223}.}
\bibitem{Bhattacharyya:2017tyc} 
  S.~Bhattacharyya and S.~Shankaranarayanan,
 \href{http://dx.doi.org/10.1103/PhysRevD.96.064044}{
  Phys.\ Rev.\ D {\bf 96}, no. 6, 064044 (2017)}
\bibitem{JMT}{H. Johansson, G. Mogull, F. Teng, \href{https://doi.org/10.1007/JHEP09(2018)080}{{\it JHEP} {\bf 09} (2018) 080}.}
\bibitem{frad}{ E. S. Fradkin and A. A. Tseytlin, \href{http://dx.doi.org/10.1016/0550-3213(82)90444-8} {{\it Nucl. Phys. B} {\bf 201} (1982) 469-491}.}

\bibitem{mali}{Mahendra Mali, Scalar-tensor theory of gravity in light-cone gauge, {\it in preparation}. }

\bibitem{ruf}{Michael S. Ruf, Christian F. Steinwachs, {\it to appear in Phys. Rev. D}, \href{http://arxiv.org/abs/arXiv:1711.07486}{ arXiv:1711.07486}.}

\bibitem{debottam}
  D.~Nandi and S.~Shankaranarayanan, \href{http://dx.doi.org/10.1088/1475-7516/2016/06/038}{JCAP {\bf 1606}, no. 06, 038 (2016)}; \href{http://dx.doi.org/10.1088/1475-7516/2016/10/008}{ JCAP {\bf 1610}, no. 10, 008 (2016)}

\bibitem{picon} 
  C.~Armendariz-Picon and G. \c{S}eng\"or,
  \href{http://dx.doi.org/10.1088/1475-7516/2016/11/016}{JCAP,  {\bf 1611}, no. 11, 016 (2016)}

\bibitem{cad}{Kasper Peeters, {\it  Introducing Cadabra: a symbolic computer algebra system for field theory problems}, \href{http://xxx.lanl.gov/abs/hep-th/0701238}{arXiv:hep-th/0701238}\\
 Kasper Peeters, {\it  A field-theory motivated approach to symbolic computer algebra}, \href{http://dx.doi.org/10.1016/j.cpc.2007.01.003}{ {\it Comput. Phys. Commun.} {\bf 176} (2007) 550-558}}


\end{thebibliography}
\end{document}